\documentstyle[11pt,newpasp,twoside]{article}
\def\edcomment#1{\iffalse\marginpar{\raggedright\sl#1\/}\else\relax\fi}
\reversemarginpar

\begin{document}

\def\kms{\hbox {\ km\ s$^{-1}\,$}}
\def\hi {H{\sc I} }
\def\hii {H{\sc II} }

\title{The W4 chimney/superbubble}
 \author{Magdalen Normandeau}
\affil{Astronomy Department, University of California, Berkeley, CA,
United States, 94720-3411}
\begin{abstract}
A conical void in the Galactic \hi within the Perseus arm has been
proposed to be a chimney. However, H$\alpha$ data
suggest that the structure may be closed at higher latitudes and
therefore is a superbubble rather than a chimney.
Recent observations have extended our view of the \hi and the radio
continuum emission to higher latitudes, up to 8\deg. The new
images show the \hi structure to be open at the top and a small
filament suggests recent breakout. The conical shape of the structure
is not easily explained by superbubble models. 
\end{abstract}

\section{The original observations: morphology and energy source}

The pilot project of the Canadian Galactic Plane Survey (CGPS) brought
to light a conical void in the \hi above the W4 \hii region
(Normandeau, Taylor, \& Dewdney 1996; hereafter NTD96). This cone is
perpendicular to the plane and opens up 
towards higher Galactic latitudes. At its apex is the OCl 352 cluster
containing nine O-stars, two of which are of very early type.  Given
this, it is unlikely that there has been a supernova in the cluster.

Within the lower density region there are \hi
``streamers'' and two elongated molecular clouds, all of which point away from
OCl 352 (Heyer et al.\ 1996 and Taylor et al.\ 1998). These
features combine to suggest that this is a Galactic chimney blown by
the stellar winds of the O-stars, with gas streaming upward toward the
halo.

\section{Is it really a chimney?}

\subsection{H$\alpha$ imaging}

Narrow band 
H$\alpha$ observations by Dennison, Topasna, \& Simonetti (1997; hereafter
DTS97) suggest the presence of 
a faint cap at $b \sim 7\deg$, corresponding to a height of
approximately 200 pc above the star cluster. Their data, scaled by
$\sin (b)$ to highlight the faint higher latitude emission, are shown
in the left-hand panel of Figure 1.

\begin{figure}[t]
\vspace{10.5cm}
\includegraphics{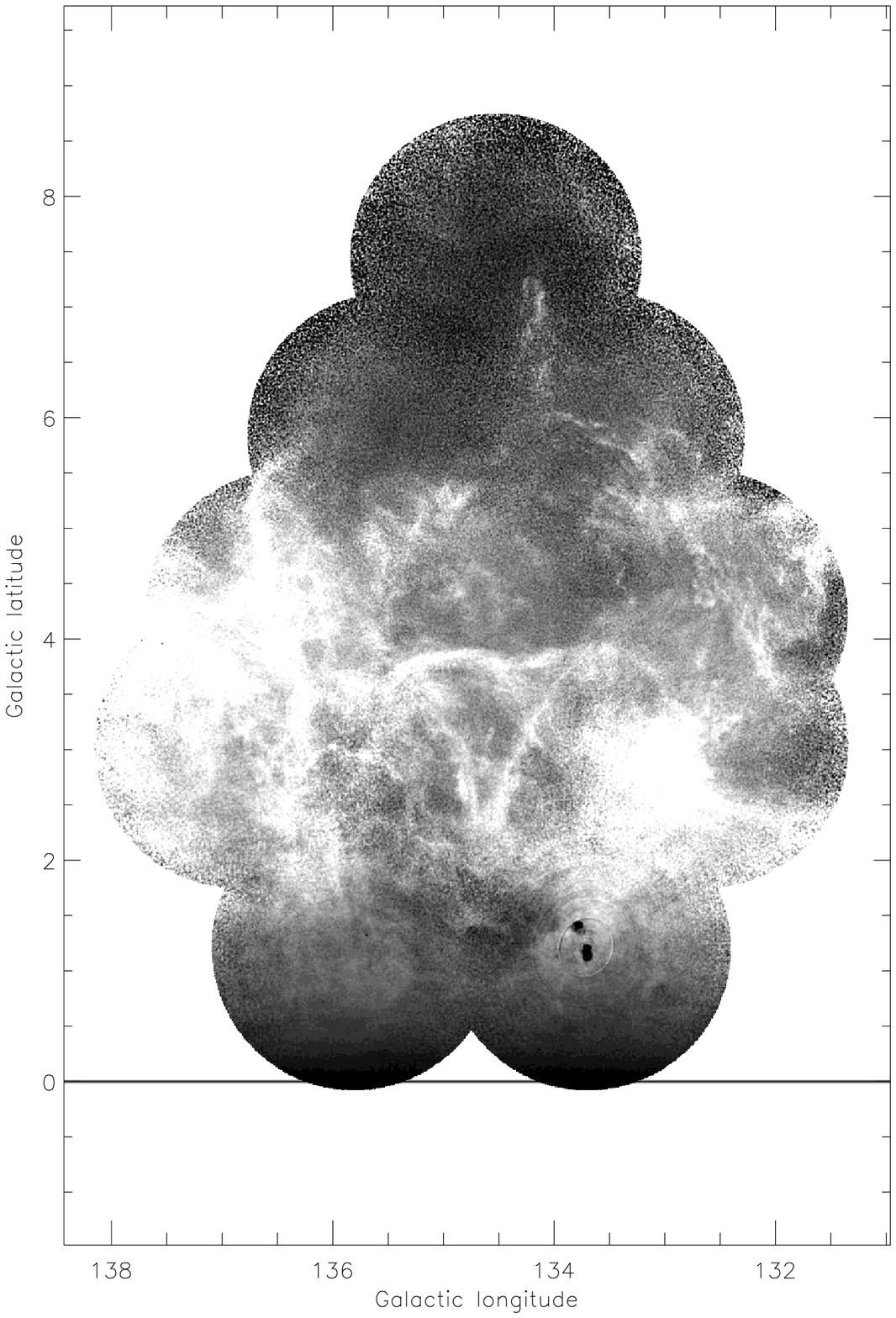}
\includegraphics{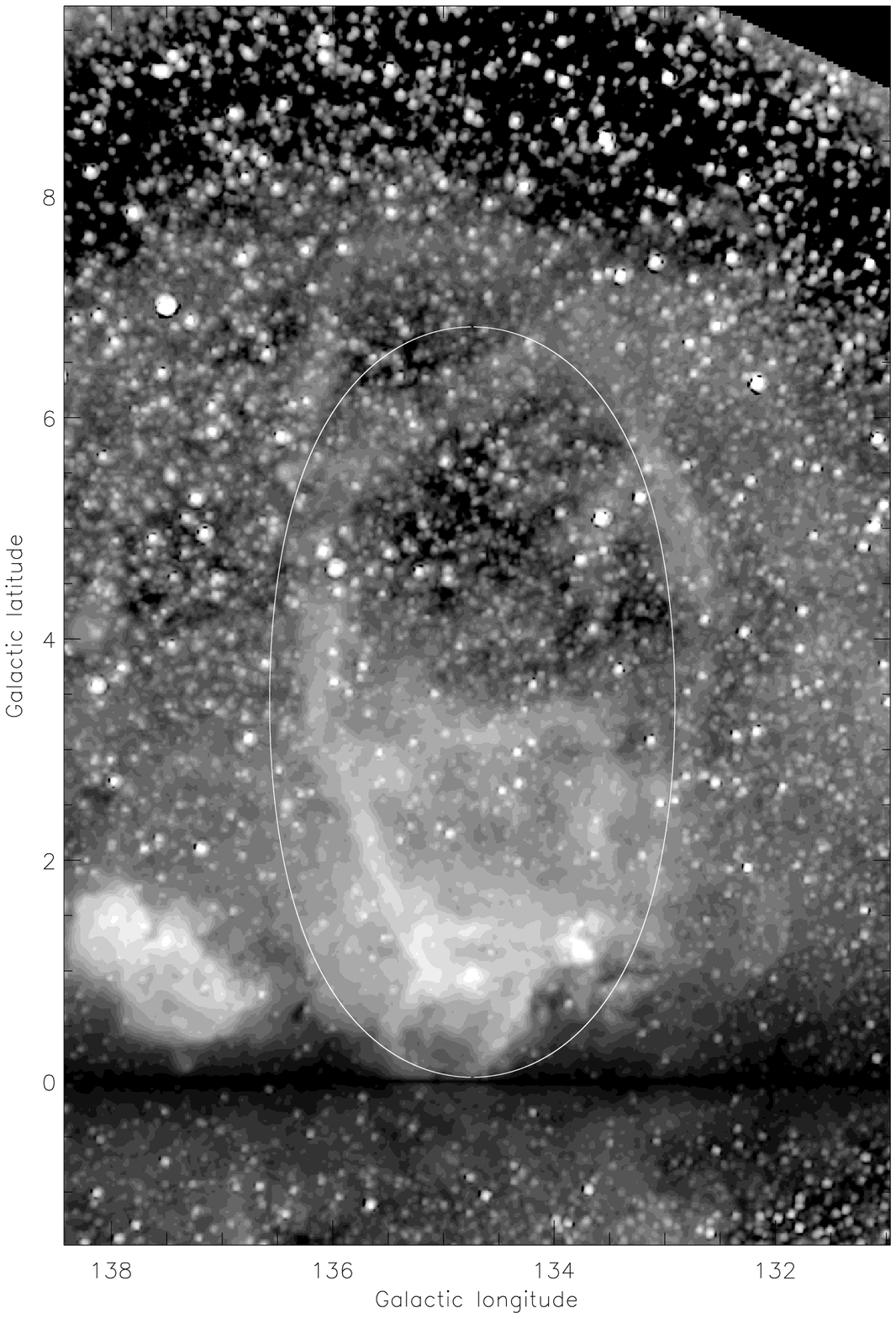}
\caption{The W4 superbubble/chimney in H$\alpha$ and \hi}
	{In both images, a $\sin (b)$ scaling has been
	introduced to highlight the low level, upper latitude
	emission, greatly surpressing the dominant Galactic plane emission. 
	{\bf Left panel}: 
	Narrow band, H$\alpha$ image (DTS97). 
	Note that the $\sim2\deg$-wide band of lower
	emission near the top of the image is an instrumental artefact. 
	The curve is the best fit Kompaneets model derived by BJM99.
	{\bf Right panel}:
	\hi image for $v_{\rm LSR} = -41.8 \kms$ to $-45.0 \kms$.}
\label{fig:Halpha_HI}
\end{figure}

\subsection{Modelling}

In NTD96, the structure's approximate age was derived from the
Weaver et al.\ (1977) formalism for a bubble expanding in a
uniform medium. However, the evolution of the superbubble must
be affected by the general decrease in density with increasing latitude.
Basu, Johnstone, \& Martin (1999; hereafter BJM99) modelled the
wind blown bubble and the ionisation structure, using both the data
from NTD96 and from DTS97. For the wind-blown bubble, they used the
Kompaneets model (Kompaneets 1960) for a shock wave propagating in an
exponential atmostphere.

The most surprising result from the dynamical modelling is
the implication of a very small scale height ($H$), namely 25 pc. This value
was obtained by matching the aspect ratio of the superbubble as seen
in H$\alpha$, assuming a distance of
2.35 kpc. The authors also point out that a small scale height is
unavoidable in any model because the current maximum radius of the
bubble must be significantly greater than $H$ in order for the bubble
to have become so elongated. 

\section{New HI data}

Data from the CGPS pilot project
(Normandeau, Taylor, \& Dewdney 1997) were combined
with observations of six new fields 
taken with the Dominion Radio Astrophysical Observatory's Synthesis
Telescope. 
The right-hand panel of Figure~1 shows the resulting mosaic for
velocities near --43.4 \kms, again scaled by $\sin (b)$ to highlight
the weaker high latitude emission.

\subsection{Scale height}

Using the
data from complementary low resolution observations, which extend
further up in latitude,  
almost to +10\deg, and fitting an exponential to the decay in
column density, a scale height of roughly 140 pc is found for
the \hi in the vicinity of the superbubble. Apart from the
contradiction with the prediction by BJM99, this result is not
particularly surprising: scale heights for the neutral medium are in
the 100-200 pc range. It does suggest that the local medium into which
the superbubble grew is quite different even from the relatively
nearby gas.

\subsection{Open or closed?}

In the new data, there is no evidence of a cap at high latitudes in
the H{\sc I}. 
The eastern \hi wall of the superbubble is clearly visible up to
5.5\deg, at which point it curves slightly inward and
disappears. The western wall is only well defined up to 3.4\deg. An open
geometry in \hi images and a closed one in data showing ionised gas
are not mutually exclusive as pointed out by BJM99. The superbubble's
shell could be sufficiently thin at high latitudes that while it
closes the shell and prevents streaming of gas towards higher
latitudes, it does not trap the ionizing radiation which then
obliterates the \hi at higher latitudes.

Extending approximately from (134.2\deg, +6.0\deg) to (134.2\deg,
+7.3\deg), 
there is a small filament of \hi pointing upward, away from the plane, 
which may be the signature of recent break out. This low level feature
is present in four channels of the mosaic, from --41.76 \kms to --46.70
\kms. It is suggestive that it is perpendicular to the tangent to and
at the (high longitude) extremity of an equally faint arc of
\hi. This arc follows the natural line that flows from the low
longitude wall to the high longitude one as seen in the \hi and seems
to mark the boundary beyond which the \hi emission is less at these
velocities. The faint arc is below the latitude of the cap claimed
by DTS97 and therefore below the upper boundary of the best fit
Kompaneets model by BJM99. The upper tip of the
vertical filament is above the latitude claimed for the ionized cap
suggested by the DTS97 data.

\subsection{Shape}

It is noticeable that the OCl 352 cluster is at the base of the
\hi cone, as was described in NTD96 whereas the Kompaneets model by
BJM99, which was fit primarily to the H$\alpha$ image, shows a bubble
extending to lower latitudes, to the base of the ionized gas loop
which forms the lower half of W4.

This cone shape with an energy source at the apex is not
unique, something similar is seen at the base
of the Aquila supershell (Maciejewski et al.\ 1996), but it is unclear
how such a structure could form. The W4 cone extends approximately 200
pc upward from the OCl 352 cluster, based on the well-defined upper
longitude wall. This is comparable to the cone at the base of the
Aquila supershell which extends roughly 175 pc for an assumed
distance of 3.3 kpc.

Tenorio-Tagle, R\'ozyczka, \& Yorke (1985) present models of
supernova remnants crossing large density discontinuities which result in
shapes that are more conical than ellipsoidal. The opening angle of the
structure above W4 is more regular however, it is more nearly a cone
than are the models.

\section{Conclusions and future endeavours}

The classification of the structure above the W4 \hii region, whether
a superbubble or a chimney, remains undecided at this time. It may be
an object in transition between the two phases of evolution. 
However, it is clear that there is no impediment to ionizing
radiation escaping towards higher latitudes and therefore the
OCl 352 cluster can contribute to maintaining the Reynolds layer of
ionized gas via this conduit.

Along with the \hi data, the DRAO observations also yield images
of the radio continuum at two frequencies, with full polarimetric
information at 1420 MHz. It is hoped that the latter data set will
help to confirm or infirm the marginal detection 
of a possible cap in the H$\alpha$ emission.

The shape of the structure along with the scale height of the adjacent
gas present challenges for modelling. The scale height inferred from
the aspect ratio is surprisingly small and may reflect a local
enhancement. The walls are very straight, making the structure conical
rather than ellipsoidal as are the results of most models. The location
of the presumed energy source at the apex of the cone rather than
within it is also unusual.


\begin{references}
\reference Basu, S., Johnstone, D., Martin, P.\ G. 1999, \apj, 516, 843
\reference Dennison, B., Topasna, G.\ A., Simonetti, J.\ H. 1997,
	\apjlett, 474, L31 
\reference Heyer, M.\ H., Brunt, C., Snell, R.\ L., Howe, J.,
	Schloerb, F.\ P., Carpenter, J.\ C., Normandeau, M., Taylor,
        A.\ R., Cao, Y., Terebey, S., Beichman, C.A. 1996, \apjlett, 464, L175
\reference Kompaneets, A.\ S. 1960, Sov.\ Phys.\ Dokl., 5, 46 
\reference Maciejewski, W., Murphy, E.\ M., Lockman, F.\ J., Savage,
        B.\ D. 1996, \apj, 469, 238 
\reference Normandeau, M., Taylor, A.\ R., Dewdney, P.\ E. 1996,
	Nature , 380, 687
\reference Normandeau, M., Taylor, A.\ R., Dewdney, P.\ E. 1997, \apjsupp,
        108, 279
\reference Taylor, A.\ R., Irwin, J.\ A., Matthews, H.\ E., Heyer, M.\
	H. 1998, \apj, 513, 339
\reference Tenorio-Tagle, G., R\'ozyczka, M., Yorke, H.\ W. 1985, \aap,
	148, 52	
\reference Weaver, R., McCray, R., Castor, J., Shapiro, P., Moore,
	R. 1977, \apj, 218, 377

\end{references}
\end{document}